# GEOGRAPHIC ROUTING PROTOCOLS FOR UNDERWATER WIRELESS SENSOR NETWORKS: A SURVEY

Sihem Souiki[1], Maghnia Feham[1], Mohamed Feham[1], Nabila Labraoui[1]

[1] STIC laboratory, Department of Telecommunication, University of Tlemcen, Algeria

**ABSTRACT**

*Underwater wireless sensor networks (UWSN), similar to the terrestrial sensor networks, have different challenges such as limited bandwidth, low battery power, defective underwater channels, and high variable propagation delay. A crucial problem in UWSN is finding an efficient route between a source and a destination. Consequently, great efforts have been made for designing efficient protocols while considering the unique characteristics of underwater communication. Several routing protocols are proposed for this issue and can be classified into geographic and non-geographic routing protocols. In this paper we focus on the geographic routing protocols. We introduce a review and comparison of different algorithms proposed recently in the literature. We also presented a novel taxonomy of these routing in which the protocols are classified into three categories (greedy, restricted directional flooding and hierarchical) according to their forwarding strategies.*

## 1. INTRODUCTION

The earth is a water planet, because more than 70% of its surface is covered by the sea and ocean, the remaining part are covered by human being. Several reasons attract to discover this underwater world such as the still large unexplored surface, the biological and geological wealth, the natural and man-made disasters, which have given rise to significant interest in monitoring oceanic environments for scientific, environmental, commercial, security and military fields [1]. Due to these reasons, underwater wireless sensor networks (UWSN) are very promising to this hostile environment. They have many potential applications, including ocean sampling networks, undersea explorations, disaster prevention, seismic monitoring, and assisted navigation [2]. The function of a routing protocol in UWSN is a fundamental part of the network infrastructure to establish routes between different nodes.UWSN routing protocols are difficult to design in general. It is a challenging task, caused by the aquatic environment. UWSN are significantly different from the terrestrial sensor technology. First, the suitable medium of communication in underwater networks is the acoustic waves and is preferred to both radio and optical waves because they have great drawbacks in aquatic channel [3]. Secondly, the most terrestrial sensors are static, while underwater sensor nodes may be mobile with water movements and other underwater activities. Consequently the challenge imposed by UWSNs leads to the inability to adapt directly the existing routing protocols in terrestrial WSN, so new routing approach must be implemented for UWSN.

In spite of the existence of a considerable number of papers about routing protocols in UWSNs presented by [4] [5] [6], we perceived a lack of a specific overview involving the geographic routing protocols. In this paper we provid e an insight into geographic routing protocols designed specifically for UWSN. In ad dition, we introduce the main challenges of using geographic routing protocols in UWSN from different perspectives and discuss some directions of future research on this field.

The remainder of this paper is structured as follows. Section 2 introduces the preliminaries, such as the architecture communication in UWSN, their challenges and basic concepts of geographic routing. While section 3 presents the classification of geographic routing protocols. Section 4 presents some details of existing routing protocols according to their classification and section 5 discusses the performance comparison of the cited protocols. Finally, we indicate in section 6 some possible future research directions and conclude the paper in section 7.

## 2. PRELIMINARIES ON UNDERWATER WIRELESS SENSOR NETWORKS

### 2.1. Architecture communication in UWSN

Similar to terrestrial sensor networks, under water sensor networks consist of a variable number of sensor nodes [7] (cabled seafloor sensors, acoustically connected sensors, moored sensors, autonomous underwater vehicle) as illustrated in Figure 1, that are deployed to perform collaborative monitoring over a given volume. The data collected by these sensors are transmitted to the surface station. The surface station is equipped with an acoustic transceiver that is able to handle multiple parallel communications with the deployed underwater sensors. It is also endowed with a long range RF and/or satellite transmitter to communicate with the onshore sink and/or to a surface sink [8].

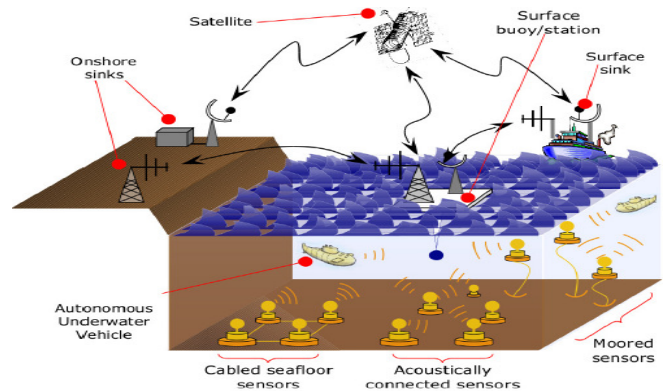

Figure 1. Different ways deployments of UWSN [7].

Underwater wireless sensor network architecture has been classified into two-dimensional and three-dimensional with fixed nodes and three-dimensional with Automatic Underwater Vehicles (AUVs) [8]. This classification is based on the geographical distribution of the nodes and their mobility. The architecture deployed depends upon the application.

#### 2.1.1. Static two-dimensional UWSNs for ocean bottom monitoring

These are constituted by sensor nodes that are anchored to the bottom of the ocean, typical applications may be environmental monitoring, or monitoring of underwater plates in tectonics.

#### 2.1.2. Static three-dimensional UWSNs for ocean column monitoring

These include networks of sensors with depth controlled by attaching each sensor node to a surface buoy, by wires of regulated length, so as to adjust the depth of each sensor node. This kind may be used for surveillance applications or monitoring of ocean phenomena (ocean bio–geochemical processes, water streams, pollution).

#### 2.1.3. Three-dimensional networks of autonomous underwater vehicles (AUVs)

These networks include fixed portions composed of anchored sensors and mobile portions constituted by autonomous vehicles.

### 2.2. Challenges of underwater wireless sensor networks

The design of underwater wireless sensor networks may be faced by several challenges [8] such as:
- Available bandwidth is severely limited.
- Underwater channel is severely impaired, especially due to multi-path and fading.
- Propagation delay in underwater is five orders of magnitude higher than in radio frequency (RF) terrestrial channels, and extremely variable.
- High bit error rates and temporary losses of connectivity (shadow zones) can be experienced, due to the extreme characteristics of the underwater channel.
- Battery power is limited and usually batteries cannot be recharged, also because solar energy cannot be exploited.
- Underwater sensors are prone to failures because of fouling and corrosion.

### 2.3. Geographic routing protocols

The major characteristic of geographic routing protocols that is involves location information in routing decisions. Location based routing is very promising for packets transmission in mobile wireless ad-hoc and sensor networks particularly in hostile environments because it does not add any burden in the network design although the localization process itself in this kind of routing is an intrinsic source of communication errors [9].Although the research on geographic routing being more recent than topological routing, it has received a special attention due to the significant improvement that geographic information can produce in routing performance. Geographic routing does not require that a node performs maintenance functions for topological information beyond its one-hop neighbourhood [10]. Consequently, geographic routing is more feasible for large-scale networks than topological routing, which requires network-wide control message dissemination. Besides that, geographic routing requires lower memory usage on nodes by maintaining the information locally [11].

The most existing geographic routing protocols adopt different policies to select the next hop. However, these policies cannot be directly applied to mobile UWSNs. First, all the existing geographic routing protocols are proposed for 2-dimensional networks; although the UWSNs are deployed in 3-dimentional environments. Second, mainly geographic routing protocols do not consider the reliability issue. They frequently adopt single forwarding path, and thus are exposed to node failure. Third, many policies are still based on relatively stable network topologies.

## 3. CLASSIFICATION OF GEOGRAPHIC ROUTING PROTOCOLS IN UWSNS

In geographic routing protocols the key information is the current position of the destination, so the sender must be aware of this important information, which can be obtained by a location service. According to how many nodes host the service four possible combinations can be resulted as some-for-some, some-for-all, all-for-some and all-for-all introduced in [12].

In this work we focus on packet-forwarding strategies as a selection criteria to introduce a novel classification of protocols. In Figure 2, geographic routing protocols are classified into three categories: greedy forwarding, restricted directional flooding and hierarchical approaches.

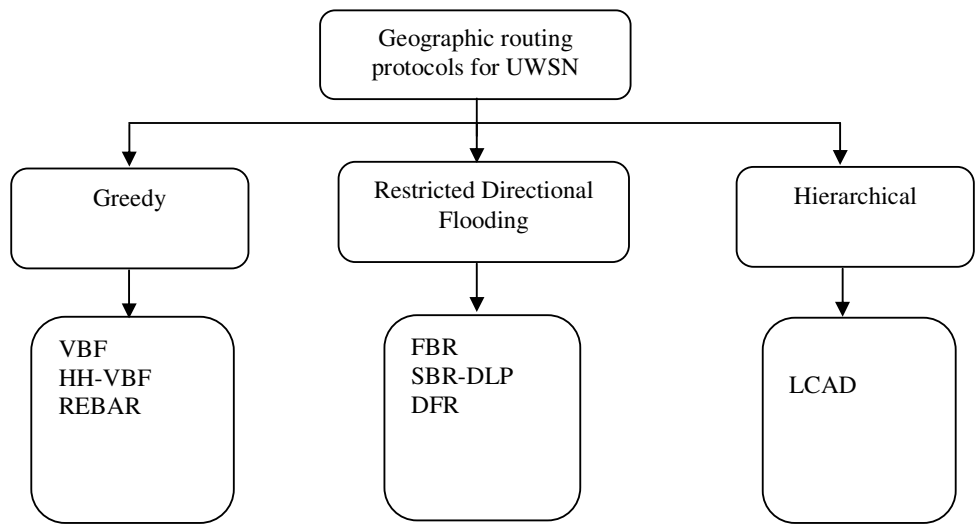

Figure 2. Classification of the geographic routing protocols for UWSN.

### 3.1. Greedy

In this category the node forwards the packet to a single node as a next hop which is located closer to the destination than the forwarding itself. Greedy protocols do not create and maintain paths from source to the destination; as an alternative, a source node includes the approximate position of the receiver in the data packet and selects the next hop according the optimization process of the protocol; the closest neighbor to the destination for example [12] [13].

To ensure the packet delivery from a source to a destination this kind of routing broadcast periodically small packets (beacons) to advertise their position and allow other nodes to maintain a one-hop neighbor table. The greedy routing can well scale with the size of network also are flexible to topology changes without using routing discovery and maintenance. However the beacons can cause a congestion problem in the network and mitigates the nodes' energy [14].

## 3.2. Restricted directional flooding

The sender will broadcast the packet (whether the data or route request packet) to all single hop neighbors towards the destination. The node which receives the packet checks whether it is within the set of nodes that should forward the packet (according to the used criteria). If yes, it will retransmit the packet. Otherwise the packet will be dropped. In restricted directional flooding, instead of selecting a single node as the next hop, several nodes participate in forwarding the packet in order to increase the probability of finding the shortest path and be robust against the failure of individual nodes and position inaccuracy.

## 3.3. Hierarchical

The third forwarding strategy is to form a hierarchy in order to scale to a large number of mobile nodes. Some strategies combine nodes' locations and hierarchical network structures by using dominating set routing such as grid in LCAD (Location-Based Clustering Algorithm for Data Gathering).

## 4. PROTOCOLS DESCRIPTION

### 4.1. Protocols based on Greedy forwarding strategy

In this section, we presented the geographic routing protocols that rely on greedy forwarding strategy.

#### 4.1.1. VBF

VBF (vector based forwarding) is the first routing protocols proposed for underwater sensor networks [15]. It is based on TBF (Trajectory based forwarding) protocols which use the source and Cartesian routing. VBF is a geographic routing protocol which requires a full localization. The position of each node is estimated with angle of arrival (AOA) technique and strength of the signal, the location information of the sender, the forwarder, and the target are carried in the packet. The path transmission is specified by a vector from a sender to a destination, and this vector is located in the center of a pipe routing, the entire nodes in this pipe are candidate for packet transmission. When a node receives a packet, it firstly calculates its position with (AOA) technique, if the node determines that it is included in the pipe, it continues transmission of the packet otherwise it discards the packet. To saving energy consumption, the selection of eligible node for packet forwarding is determinate with a desirableness factor which is defined as:

$$\alpha = \frac{p}{W} + \frac{(R - d \times \cos\vartheta)}{R} \quad (1)$$

Where p is the projection of A to the routing pipe $\overrightarrow{S1S0}$, d is the distance between the candidate forwarding node A and the current forwarding node F, $\vartheta$ is the angle between the vector $\overrightarrow{FS0}$ and $\overrightarrow{FA}$, R is the transmission range, w is the radius of route pipe.

After calculating the desirableness factor the node holds this packet for a time period $T_{adaptation}$ which is defined as:

$$T_{adaptation} = \sqrt{\alpha} \times T_{delay} + \frac{R-d}{v0} \quad (2)$$

Where $T_{delay}$ is a pre-defined maximum delay, called maximum delay window and      is the propagation speed of acoustic signals in water (1500 m/s), and d is the distance between this node and the forwarder. During the $T_{adaptation}$, if a duplicated packet is received from different node, the node compares its desirableness factor with other node and decides about the forwarder of the packet.

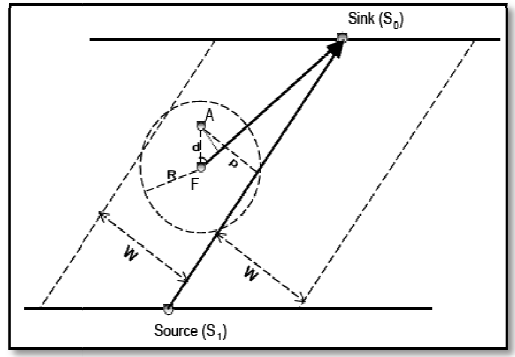

Figure 3. Desirableness factor of VBF routing protocol

**4.1.2. HH-VBF**

The performance of VBF protocol can be decreased on account of two fundamental problems. The first is the sensitivity to the routing pipe's radius and the second is the low delivery ratio in sparse networks.

To overcome the drawbacks of VBF, the hop by hop VBF (HH-VBF) protocol [16], HH-VBF shares some characteristics of VBF protocol such as geographic and source routing. In VBF routing protocol a unique virtual pipe is created from the source to the sink, however in HH-VBF at each hop a virtual pipe routing is created, so a hop-by-hop approach is used in the routing operation.

Upon receiving a packet from a source or a forwarder, the node computes the vector from its sender toward the sink, and then it calculates its distance to that vector. If this distance is smaller than a radius of the virtual routing pipe, this node is qualified for the transmission of packet and becomes a candidate forwarder.

The HH-VBF protocol uses a self adaptation algorithm but in different way as in VBF, the desirableness factor is defined by the following equation:

$$\acute{\alpha} = \frac{(R - d \times \cos\theta)}{R} \qquad (3)$$

After calculating this factor the packet will be holding for a time period $T_{adaptation}$ as in VBF protocol. The suppression strategy of duplicate packets forwarding is handled by overhearing the transmission of the same packet multiples times in the network. If a node receives a duplicate packet, its calculates its distance from each neighboring nodes which are forwarding the packet to the sink.

If the small distance among these distances is still larger than a predefined threshold, the node transmits the packet; otherwise the packet is dropped [5].

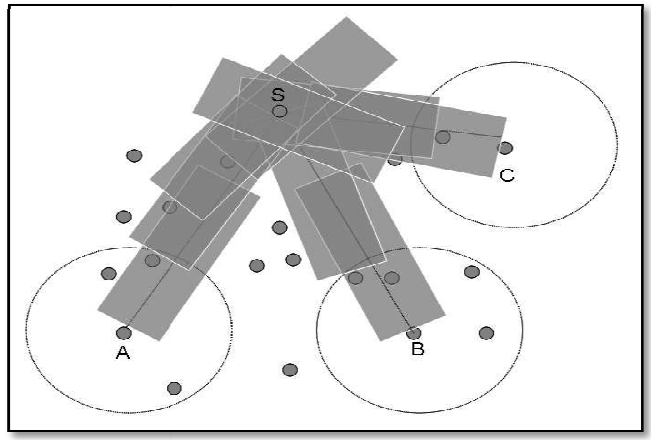

Figure 4. HH-VBF routing protocol

### 4.1.3. REBAR

The proposed protocol reliable and energy balanced algorithm routing (REBAR) [17], is a location based routing protocol that focuses on three significant problems to deal in UWSNs: energy consumption, delivery ratio and handling void problem. First, REBAR uses a sphere energy depletion model to analyze the energy consumption of nodes in UWSNs. Then this model is extended by considering the node mobility in UWSNs, and they assumed that node mobility is a positive factor which can help balance the energy depletion in the network and prolong lifetime of networks. In REBAR, nodes broadcast in a specific domain between source and sink using geographic information since network-wide broadcast causes high energy consumption. Thus, the size of the broadcast domain is critical. Consequently an adaptive scheme is designed for setting broadcast domain size. In particular, the constrained radius of nodes is set to different values depending on the distance between the nodes and the sink. Nodes nearer the sink are set to smaller value in order to reduce the chance of being involved in the routing, thus balancing the energy consumption among the nodes.

The routing process of REBAR consists that each node in the network has a constrained radius which is concerned with its distance to sink. The source calculates a directional vector v from itself to destination. The Euclidian distance from source to sink d and the vector v are stored in the packet. The packet is assigned with a unique identifier (ID), which is composed of the source ID and a sequence number. The packet is broadcasted in the network. Each receiver maintains a buffer to record the ID of recently received packets. Duplicates can be treated by the history and will be discarded. In order to ensure that the packets are forwarded towards the sink, the following scheme is adopted. When a neighbouring node i of the source node receives a packet for the first time, it first compares its distance $d_i$ to sink with d. It drops the packet if
$(d_i - d)$ is greater than a threshold. This comparison ensures that packets are transmitted in the right direction. If the calculated distance to the vector v by the receiver is larger than its constrained radius, the packet is dropped. Otherwise, the receiver forwards the packet. By this way, the broadcast is constrained in a reasonable domain, and packets are delivered in redundant and interleaved paths. Figure 5 (b) depicts the illustration of the routing process of REBAR.

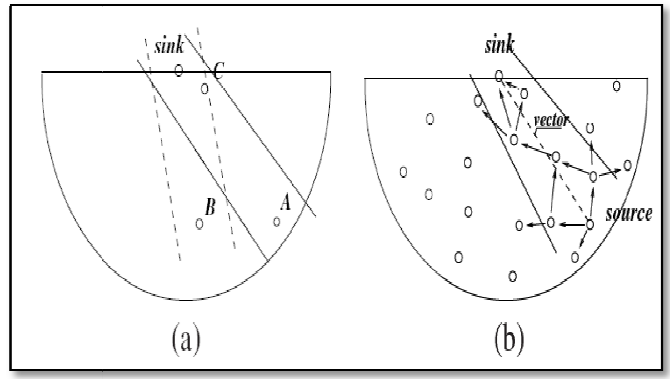

Figure 5. The routing process of REBAR

REBAR uses an extended mechanism to bypass routing voids in the network by the assumption that the nodes on the boundary of the voids can detect the existence of voids using methods proposed in [18] [19]. The nodes in the network can be divided into two different sets: Boundary-Set and Non-Boundary-Set. When a node in the Non-Boundary-Set receives a packet, it behaves as in the basic REBAR. While a node in the Boundary-Set, forwards the packet to all their neighbors directly without checking distance and vector information.

## 4.2. Protocols based on restricted directional flooding forwarding strategy

### 4.2.1. FBR

The focused beam routing (FBR), is a location based and energy efficient routing protocols. FBR assumes that each node knows only its own location and the final destination location. In the proposed protocol, variable transmission power levels are used in the forwarding of data packet, and this transmission power have a range from $P_1$ to $P_n$. For each power level there is a corresponding transmission radius $d_n$, which is a cone of angle emanating from the source node to the destination [20].

The selection of the next forwarder is done as follow:

Firstly, source node multicasts an RTS in their neighborhood with the lower power level $P_1$, in the second step we have three cases:

**1/ one reply with CTS packet:** If only one node exist in the transmission radius, it will reply by a CTS and will be the forwarder
**2/ multiple replies with CTS packet:** In this case the source node chooses the most desirable node (the closest to the destination) for the transmission of packet.
**3/ any CTS packet are replayed:** If there is no CTS packet received, so the power level must be increased to the higher level until receiving a CTS reply. If the maximum level is reached without receiving a CTS packet thus the cone of angle must be shifting in the left or the right of the first cone.

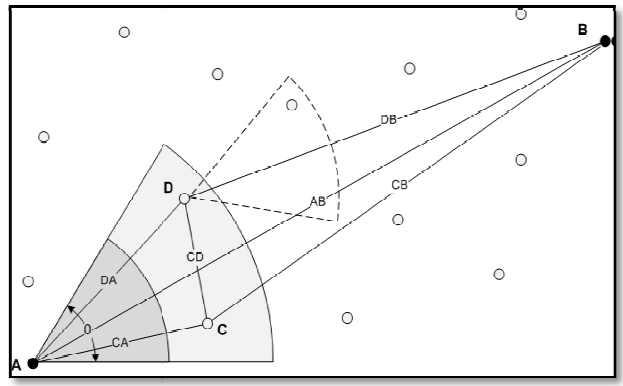

Figure 6. Illustration of the routing protocol: nodes within the transmitter's cone θ are candidate relays.

**4.2.2. DFR**

The high mobility of nodes and the conditions of aquatic environment are two factors that lead to the packets loss in routing processes and that decrease the reliability of networks the proposed directional flooding based routing protocol (DFR), that focuses on these problems and takes into account the link quality in the forwarding strategy [21].

The proposed approach, assumed that the geographic information is available; all nodes know their location, their one-hop neighbors' location and the location of a sink. In addition, the nodes are also able to measure the quality of the links with neighbors. DFR also addresses a well-known void problem by allowing at least one node to participate in forwarding a packet. In DFR, a packet transmission uses a scoped flooding, i.e. a flooding zone is created to limit the flooding in the whole networks. This flooding zone is determined using an angle between      and     , where F is the node which receives a packet, S and D are the source and destination, respectively.

Upon receiving a data packet, F determines the packet forwarding by comparing the angle (SFD) with an angle for flooding, called BASE_ANGLE, which is involved in the received packet. The BASE_ANGLE is initially set to a value A_MIN (a threshold). The node upon receiving the broadcast packet computes an angle (CURRENT_ANGLE) among a source, itself and a sink. If the node's CURRENT_ANGLE is smaller than the BASE_ANGLE, the node discards the packet because it is considered out of the flooding scope. Otherwise, the receiving node adjusts the BASE_ANGLE according to the link quality of its neighbors and transmits the packet.

The BASE_ANGLE can be adjusted on the basis of the link quality by modifying the size of the flooding zone, When the average link quality is worse than a threshold, a node decreases the BASE ANGLE by a predefined decrement value (A_DCR) and forwards the packet, otherwise the node increases the BASE ANGLE by a predefined increment value (A_ICR) and forwards the packet.

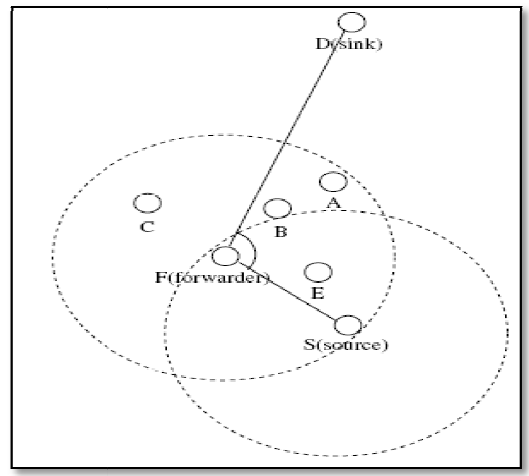

Figure 7. An example of a packet transmission in DFR

### 4.2.3. SBR-DLP

Generally the proposed location based routing protocols for underwater sensor networks assumes that the locations of destination nodes are frequently fixed but according to the random motion by ocean currents and the self-propelling capability of node, such assumptions are usually invalid in underwater sensor networks.

A sector-based routing with destination location prediction (SBR-DLP) have introduced in [22], it supposes that a node knows its own location, and the location of the destination node is predicted. Consequently, it relaxes the need for accurate knowledge of the destination's location.

In SBR-DLP the sensor nodes are not required to carry neighbor information or network topology. Each node is assumed to know its own position, and the destination node's pre-planned movements. This movement is usually predefined prior to launching the network.
A hop-by-hop fashion is used to route the packet to the destination, instead of finding the complete path before sending a packet. As shown in Figure 8, a node S has a data packet that needs to be sent to destination D. Firstly, it will try to find its next hop by broadcasting a Chk_Ngb packet, which includes its current position and packet ID. The neighbor node that receives Chk_Ngb will check whether it is nearer to the destination node D than the distance between nodes S and D. The nodes that meet this condition will reply to node S by sending a Chk_Ngb_Reply packet.

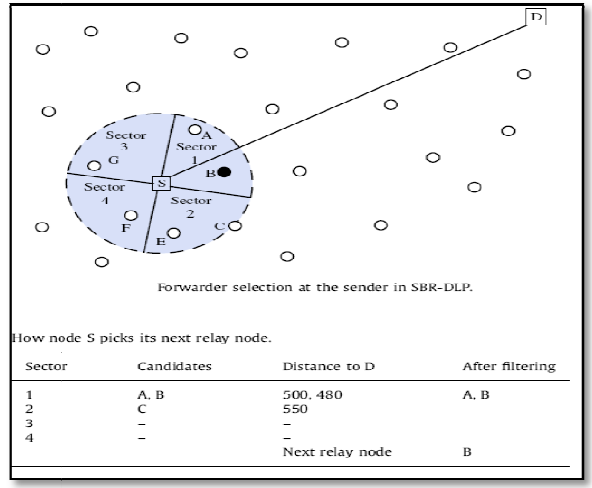

Figure 8. The SBR-DLP routing protocol.

The SBR-DLP allows the sender determine its next hop using information received from the candidate nodes. Thus proposed SBR-DLP is different from both VBF and HH-VBF; which let each candidate node decide whether it should relay the packet; this eliminates the problem of having multiple nodes acting as relay nodes, which is encountered in both VBF and HH-VBF.

Although the SBR-DLP shares some similarities with the FBR (e.g., letting the sender decide its next relay node), there are some important differences. FBR uses a single transmitting cone that covers only a part of the communication area, the SBR-DLP considers the whole communication circle to locate the candidate relay nodes. In addition, while the FBR needs to rebroadcast the RTS every time it cannot find a candidate node within its transmitting cone, the SBR-DLP does not need to do so.

### 4.3. Protocols based on hierarchical forwarding strategy

#### 4.3.1. LCAD

Several clustering Algorithms such as LEACH [23], HEED [24], and PEGASIS [25] have been proposed for terrestrial sensor networks. These protocols cannot be directly applied to underwater sensor networks due to the nature of the aqueous media. The underwater channel uses the acoustic waves and hence the propagation delay incurred in an underwater sensor network is higher than its terrestrial counterpart. Moreover the sensors are deployed in a 3-Dimensional topology.

A clustering algorithm based on the geographical location of the sensor nodes in 3-D Hierarchical network architecture called LCAD have proposed in [26].

In this protocol, the entire network is divided into 3-dimensional grids. The architecture of the network is shown in Figure 9. The optimal horizontal transmission range is less than 50m and the vertical transmission range is around 500m; the size of each grid is set approximately to 30m x 40m x 500m. A grid comprises of a single cluster.

The data communication is composed of three phases: (i) set-up phase, where the cluster head is selected. (ii) Data gathering phase, where data is sent by the nodes in the cluster to the cluster head. (iii) Transmission phase, where the data gathered by the cluster heads is transmitted to the base station.

International Journal of Wireless & Mobile Networks (IJWMN) Vol. 6, No. 1, February 2014

Some of the sensor nodes in a cluster have additional capabilities in terms of memory and energy. Such nodes are qualified as cluster heads (ch). Having multiple ch-nodes guarantees reliability and load balancing in the network. These ch-nodes are located approximately in the centre of the grid, thereby enabling them to communicate with a maximum number of non-cluster nodes, within the grid in an energy efficient manner. The grids are organized in a manner similar to the cells in a cellular network. While a cell has a single fixed base-station, a grid has multiple ch-nodes and the role of cluster head is rotated amongst these.

The selection of the cluster head is based on the sleep wake pattern along with residual memory and energy of the contending ch-nodes.
LCAD uses two-level addressing scheme within the network, the first is used for intra-cluster communication, and the second for inter-cluster communication, using a 32-bit address format (similar to the IPv4 format).
 In intra-cluster communication the format of the address used is: **GRID.X.Y.Z** Where GRID gives the number of the grid in which the nodes are residing. X, Y and Z depict to the relative X, Y, Z position of the nodes in the grid. IDs beyond **255.0.0.0** are reserved for ch-nodes and they are used for inter-cluster communication
For inter-cluster communication the format of the address is: **255.GRID. X.Y.** The Z-position is not used for obtaining the address because we have ensured that the cluster heads are deployed at the vertical centre of each grid. Hence the Z-position is required only for intra-cluster communication. With this addressing scheme we have 256x256x256 unique addresses in a
grid. Hence the density of the sensor deployment is greater than 27 nodes/m3.

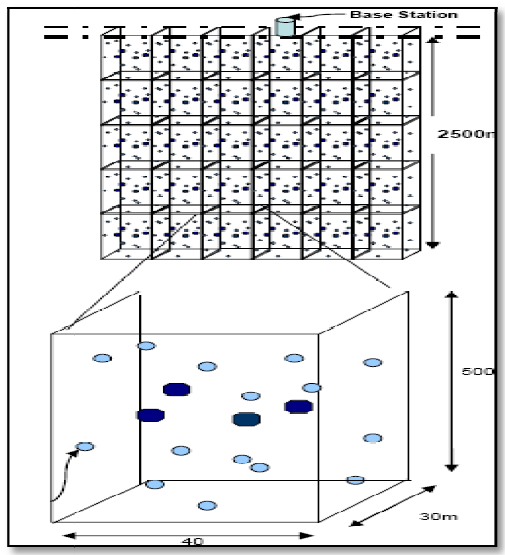

Figure 9. Architecture used in LCAD protocol with the projection of a single grid.

## 5. COMPARISON STUDY

In this section, we attempt to compare between the selected geographic routing protocols, reviewed in the last section. Therefore, these protocols are compared in a number of different ways: *forwarding strategy* (type, shape region, robustness, scalability, and packet overhead), *location service* (type, robustness), *design goal* (density, mobility, void handling, and destination mobility). We summarize this comparison in Table 1. A brief explanation for these metrics follows:

| protocol | Forwarding Strategy | | | | | Location Service | | Design Goal | | | |
|---|---|---|---|---|---|---|---|---|---|---|---|
| | Type | Shape Region | Robustness | Scalability | Packet Overhead | Type | Robustness | Density | Mobility | Void Handling | Destination Mobility |
| VBF | Greedy | Single Pipe routing | High | High | Low | All-for-some | Medium | Dense | Both | No | No |
| HH-VBF | Greedy | Per-hop pipe routing | High | High | Low | All-for-some | Medium | sparse | Both | No | No |
| REBAR | Greedy | Specific domain | High | High | Low | All-for-some | Medium | Dense | mobile | Yes | No |
| FBR | RDF | Cone | Medium | Medium | Medium | All-for-some | Medium | Sparse | Both | No | No |
| DFR | RDF | Base angle | Medium | Medium | Low | All-for-some | Medium | Sparse | mobile | Yes | No |
| SBR-DLP | RDF | sector | Medium | Medium | Medium | -- | -- | Sparse | mobile | No | yes |
| LCAD | Hierarchical | Grid routing | Medium | High | Low | -- | -- | Dense | mobile | No | No |

Table 1. Comparison of the geographic routing protocols designed for UWSN.

### 5.1. Forwarding strategy

- **Type**

As we have seen in section 3 we can classify the protocols on three basic strategies used for packet forwarding: greedy (VBF, HH-VBF, REBAR), restricted directional flooding (FBR, DFR, and SBR-DLP) and hierarchical (LCAD).

- **Shape region**

In order to minimize the energy consumption each protocols aims to limit the number of candidates relay that are qualified by the packet transmission. These protocols used different shape for this purpose, for example in VBF and HH-VBF a pipe routing is used but in HH-VBF a pipe routing is created in each hop, also REBAR uses a specific domain. In case of FBR the forwarders are restricted in a transmitting cone.

In SBR-DLP, a circle is divided on several sectors which may contain the transmitting node. In DFR only the nodes belonging to a BASE_ANGLE are involved in transmission operation, finally LCAD, which is a hierarchical protocol uses grid by grid routing.

- **Robustness**

The robustness of an approach is considered to be high if the failure (or absence due to mobility) of a single intermediate node does not prevent the packet from reaching its destination. It is the case in VBF, HH-VBF, and REBAR, we find that VBF is robust against packet loss and node failure in that VBF uses redundant paths to forward the data packets. Some of these paths are interleaved, some are parallel. However HH-VBF is more robust than VBF especially in sparser networks, it can find more paths for data delivery compared to VBF, by using the hop-by-hop vector for packet forwarding. Similar to VBF, REBAR robustness is high since the packets are delivered in redundant and interleaved paths.

In FBR, DFR, SBR-DLP and LCAD, The robustness is considered to be medium, owing to the failure of a single intermediate node might lead to the loss of the packet but does not require the set up of a new route. However, especially in sparser networks the robustness can be degraded and the setting up of a new route is required such as in FBR and DFR. In FBR the transmitting node will keep increasing the power until it reaches someone, or until all power levels have been exhausted. If it cannot reach anyone at the maximal level PN, the transmitter will shift its cone and start looking for new candidate relays left and right of the main cone. The problem still encountered in DFR, when there are no nodes closer to the sink.

- **Scalability**

We can determine the scalability performance of the protocol with an increasing number of nodes in the network. It can be classified as follows: high scalability, when a network grows as much as it needs and the approach is still able to maintain a good performance. As the case of the three greedy routing protocols VBF, HH-VBF, and REBAR because they do not need routing discovery and maintenance [27]. Moreover, they have a low packet overhead due to the small number of small-size packets and reduction of the use of control messages. LCAD uses a clustering approach which is a favorite to large scale networks. The rest of protocols have a medium scalability because that can handle networks with a reasonable size, but may have problems if it grows. Since all the position-based routing protocols are scalable compared to topology-based ones, all the discussed protocols have at least medium scalability.

- **Packet overhead**

A higher number of signaling packets and large packets' sizes lead to bandwidth consumption. Since all the discussed protocols are considered to have small packets, compared to secure protocols for example. Note that position-based routing protocols have lower packet overhead compared to topology-based ones. For example, in LCAD, the size of a control packet (CHADV, CHJOIN, D-START) is fixed to 128 bits while the size of a data packet is fixed to be 128 bytes. Hence all the discussed protocols have at most medium packet overhead.

## 5.2. Location service

- **Type**

Indicates the type of the location service used with the given protocol. It shows how many nodes participate in providing location information and for how many other nodes each of these nodes maintains location information.

VBF, HH-VBF, REBAR, FBR, and DFR use all-for-some location service; so nodes know their location, their one-hop neighbors' location and the location of a sink.

- **Robustness**

It is considered to be low, medium or high depending on whether the position of a given node will be inaccessible upon the failure of a single node, the failure of a small subset of the nodes or the failure of all nodes, respectively.

Hence, in the proposed protocols, a given node will be inaccessible upon the failure of a subset of nodes. Thus their location services robustness is regarded to be medium.

### 5.3. Goal design

- **Density**

Indicates whether the protocol is more suitable to be implemented in dense or/and sparse networks. VBF is suitable for dense networks because the packet delivery ratio is decreased for sparse networks whereas it is increased in dense networks. On the other side HH-VBF is more favorable for sparse networks, because it has a good delivery ratio in this kind of networks. Its can find more path towards destination when the density of networks is low, in addition increasing node density in HH-VBF brings a high energy cost. In spite of the medium packet overhead added by FBR, SBR-DLP and DFR, we notice that are simulated for networks with reasonable size. REBAR and LCAD are appropriate for dense networks, for instance LCAD protocol apply an addressing scheme with density of the network greater than 27 nodes/m3.

- **Mobility**

Indicates whether protocols used for mobile/static networks or both. We notice that VBF, HH-VBF and FBR can be applied within both mobile and static networks. Although the rest of routing protocols are designed for mobile networks on account of high mobility node imposed by ocean currents.

- **Void handling**

In the realistic scenarios, some regions may be uncovered by the network due to underwater obstacles or node failures. We notice that all the proposed routing protocols are designed without addressing the void problem except REBAR and DFR.

- **Destination Mobility**

In the all proposed protocols, SBR-DLP is the only one which relaxes the need for precise knowledge of the destination's location. It predicts the location of the destination node, by assuming that its pre-planned movements (its waypoints and their corresponding schedule) are made known to all other nodes before launching. However, it is important to note that the destination node can deviate from its schedule due to the ocean currents.

### 6. FUTURES RESEARCH DIRECTIONS

According to the comparison and discussion of the geographic routing protocols for UWSNs in Section V, there exist open issues which are worth focusing on:

- **Localization problem**

Generally the routing decision in geographic routing protocols is based on the destination's position contained in the packet and the position of the forwarding node's neighbors. Thereby the awareness of the location of a node can enhance the performance of the network and add significance to the information that is gathered.

Localization in underwater sensor networks is a hot topic research; hence research should consider the development of new and robust location services and should investigate the impact of various localization techniques on the performance of the geographic routing algorithms. Consequently the relationship between localization and geographic routing is proportional, improving localization leads to increase the performance of networking and several UWSN applications.

- **Void problem**

The void problem is addressed by several studies in terrestrial sensor networks which aimed the stationary and two-dimensional wireless networks. However these techniques are not suitable for underwater sensor networks because the underwater void is characterized as three dimensional spaces. In addition, the mobility of most underwater nodes makes the void mobile that can also result from the surrounding environment [28]. For example, when a ship navigates over the underwater sensor network, it blocks communications in the nearby area and thus generates a void that moves along with the ship. The characteristics of underwater sensor networks make it more difficult to manage the three-dimensional and mobile voids in such networks. Only a few geographic routing protocols take in account the void problem in their design, so we should give more importance to this challenging problem.

- **Security**

The attacks against geographic routing in UWSNs are the same as in terrestrial sensor networks. The same countermeasures cannot be directly applied to UWSNs due to their difference in characteristics such as: the large propagation delays, the low bandwidth, the difficulty of recharging batteries of underwater sensors, and the high mobility of nodes.

A lot of effort was already put in securing traditional WSN presented in [29] [30]. The security research for UWSN routing and especially position-based routing is still in its infancy. In Dis-VoW a wormhole attack can still be hidden by falsifying the buffering times of distance estimation packets [31]. The wormhole-resilient neighbor discovery presented in [32] is affected by the orientation error between sensors [33].

When using position-based routing, the most important aspect is the correctness of position data when false position information is distributed in the UWSN. This can seriously affect the performance of the network [34]. Particularly due to the fact that most protocols broadcast position information in the clear, allowing anyone within range to receive it. Therefore, node position can be falsified, making other nodes believe that it is in a different position. The nodes may believe that the malicious node is the closest to the destination and choose it as the next hop. Thus, this attacker will be able to modify or drop packets [35]. Consequently, there is a need to develop new techniques against several attacks from malicious and compromised nodes. In addition we must focuses on the location privacy which is one of the most major challenges to be tackled.

- **Energy consumption**

Energy consumption is a crucial factor to determine the life of a sensor network because generally sensor nodes are powered by battery, so the algorithm should guarantee QoS while taking into account the limited power of nodes, and scalability with network size. The challenge in UWSN is then to improve a location based routing protocol that can meet these requirements while reducing compromise.

## 7. CONCLUSION

The design of any routing protocol depends on a specific goals and requirements. Development of a geographic routing protocol for the aquatic environments is regarded as a vital research area, which will make these networks much more reliable and efficient. In this paper we have conducted a comprehensive survey of various geographic routing protocols in underwater wireless sensors networks. We classified the geographic routing protocols according to their forwarding strategies into three categories: greedy, restricted directional flooding and hierarchical approaches. We presented a performance comparison of the most relevant routing protocols in terms of forwarding strategy (type, shape region, robustness, scalability, packet overhead), location service (type, robustness), design goal (density, mobility, handling void and destination mobility).

One of the future goals in designing geographic routing algorithms is adding security mechanisms, and enhancing energy consumption of the networks.

The detailed descriptions of the selected protocols contribute in understanding the direction of the current research on location based routing protocols for UWSN and its benefits look very promising for the future networks design.